\documentclass[12pt]{article}

\def\stc{{\rm sc}}

\usepackage{amsmath}
\usepackage{float}
\usepackage{amsthm}
\usepackage{amsfonts}
\usepackage{amstext}
\usepackage{graphics}
\usepackage{epsfig}
\usepackage{fullpage}

\def \mod#1 #2{#1\ ({\rm mod}\ #2)}

\newtheorem{theorem}{Theorem}
\newtheorem{lemma}[theorem]{Lemma}
\newtheorem{corollary}[theorem]{Corollary}
\newtheorem{proposition}[theorem]{Proposition}

\newtheorem{examp}[theorem]{Example}
\newenvironment{example}{\begin{examp}\normalfont\quad}{\end{examp}}
\newtheorem{openp}[theorem]{Open Problem}

\def \intersect{\ \cap \ }
\def \union{\ \cup \ }
\def \sc{{\rm sc}}
\def \nsc{{\rm nsc}}

\begin{document}
\title{The Frobenius Problem in a Free Monoid}
\author{
Jui-Yi Kao,
Jeffrey Shallit, and Zhi Xu\\
School of Computer Science\\
University of Waterloo\\
Waterloo, Ontario N2L 3G1\\
CANADA \\
{\tt  Eric.Kao@tdsecurities.com} \\
{\tt shallit@graceland.uwaterloo.ca} \\
{\tt z5xu@cs.uwaterloo.ca} \\
}

\maketitle

\begin{abstract}
      The classical Frobenius problem is to compute the largest number
$g$ not representable as a non-negative integer linear combination of 
non-negative integers
$x_1, x_2, \ldots, x_k$, where $\gcd(x_1, x_2, \ldots, x_k) = 1$.  In this
paper we consider generalizations of the Frobenius problem to the 
noncommutative setting of a free monoid.  Unlike the commutative case,
where the bound on $g$ is quadratic, we are able to show exponential or
subexponential behavior for an analogue of $g$, depending on the 
particular measure chosen.
\end{abstract}

\section{Introduction}
\label{intro}

      Let $x_1, x_2, \ldots, x_k$ be positive integers.
It is well-known that every sufficiently large integer can be
written as a non-negative integer linear combination of the $x_i$
if and only if $\gcd(x_1, x_2, \ldots, x_k) = 1$.

      The {\it Frobenius problem} (so-called because, according to
Brauer \cite{Brauer:1942}, ``Frobenius mentioned it occasionally in his
lectures'') is the following:

      Given positive integers $x_1, x_2, \ldots, x_k$ with
$\gcd(x_1, x_2, \ldots, x_k) = 1$, find the largest positive integer
$g(x_1, x_2, \ldots, x_k)$ which {\it cannot} be represented as a 
non-negative integer linear combination of the $x_i$.

\begin{example}
{\it The Chicken McNuggets Problem} 
(\cite[pp.\ 19-20, 233--234]{Vardi:1991}, \cite{Owens:2003}).
If Chicken McNuggets can be
purchased at McDonald's only in quantities of $6$, $9$, or $20$ pieces,
what is the largest number of McNuggets that cannot be purchased?
The answer is $g(6,9,20) = 43$.
\end{example}

      Although it seems simple at first glance, the Frobenius problem
on positive integers has many subtle and intriguing aspects that continue
to elicit study.  A recent book by 
Ram{\'\i}rez Alfons{\'\i}n \cite{Ramirez-Alfonsin:2005}
lists over 400 references on this problem.
Applications to many different fields exist:  to algebra \cite{Kunz:1970};
the theory of matrices \cite{Dulmage&Mendelsohn:1964},
counting points in polytopes \cite{Beck&Diaz&Robins:2002};
the problem of efficient sorting using Shellsort
\cite{Incerpi&Sedgewick:1985,Sedgewick:1986,Weiss&Sedgewick&Hentschel&Pelin:1988,Selmer:1989};
the theory of 
Petri nets \cite{Teruel&Chrzastowski-Wachtel&Colom&Silva:1992};
the liveness of weighted circuits \cite{Chrzastowski-Wachtel&Raczunas:1993};
etc.

      Generally speaking, research on the Frobenius problem can be
classified into three different areas:

\begin{itemize}

\item Formulas or algorithms for the
exact computation of $g(x_1, \ldots, x_k)$, including formulas for $g$
where the $x_i$ obey certain
relations, such as being in arithmetic progression;

\item The computational complexity of the problem;

\item Good upper or lower bounds on $g(x_1, \ldots, x_k)$.  

\end{itemize}

      For $k = 2$, it is folklore that 
\begin{equation}
g(x_1, x_2) = x_1 x_2 - x_1 - x_2;
\label{g2}
\end{equation}
this formula is often attributed to Sylvester \cite{Sylvester:1884},
although he did not actually state it.  Eq.~(\ref{g2})
gives an efficient algorithm to compute $g$ for two elements.
For $k = 3$, efficient algorithms have been given by
Greenberg \cite{Greenberg:1988} and
Davison \cite{Davison:1994}; if $x_1 < x_2 < x_3$, these algorithms
run in time
bounded by a polynomial in $\log x_3$.    Kannan
\cite{Kannan:1989,Kannan:1992} gave a very complicated algorithm that runs in
polynomial time in $\log x_k$ if $k$ is fixed, but is wildly exponential
in $k$.  However, Ram{\'\i}rez
Alfons{\'\i}n \cite{Ramirez-Alfonsin:1996} proved that the general problem is
NP-hard, under Turing reductions, by reducing from the integer
knapsack problem.  So it seems very likely that there
is no simple formula for computing $g(x_1, x_2, \ldots, x_k)$ for
arbitrary $k$.  Nevertheless, recent work by Einstein, Lichtblau,
Strzebonski, and Wagon \cite{Einstein&Lichtblau&Strzebonski&Wagon:2007}
shows that in practice the Frobenius number can be computed relatively
efficiently, even for very large numbers, at least for $k \leq 8$.

    Another active area of interest is estimating how big $g$ is in terms
of $x_1, x_2, \ldots, x_k$ for $x_1 < x_2 < \cdots < x_k$.
It is known, for example, that
$g(x_1, x_2, \ldots, x_k) < x_k^2$.  This follows from
Wilf's algorithm \cite{Wilf:1978}.  Many other bounds are known.

     One can also study variations on the Frobenius problem.  
For example,
given positive integers $x_1, x_2, \ldots, x_k$ with
$\gcd(x_1, x_2, \ldots, x_k) = 1$, what is the number $f(x_1, x_2, \ldots, x_k)$
of positive integers not represented as a non-negative integer linear
combination of the $x_i$? Sylvester, in an 1884 paper \cite{Sylvester:1884}, 
showed that $f(x_1, x_2) = {1 \over 2} (x_1 - 1)(x_2 -1)$.

      Our goal in this paper is to generalize the Frobenius problem
to the setting of a free monoid.    In this framework, we start with
a finite, nonempty alphabet $\Sigma$, and consider the set of all finite
words $\Sigma^*$.
Instead of considering integers $x_1, x_2, \ldots, x_k$, we consider words
$x_1, x_2, \ldots, x_k \in \Sigma^*$.  Instead of considering linear
combinations of integers, we instead consider the languages
$\lbrace x_1, x_2, \ldots, x_k \rbrace^*$ and
$x_1^* x_2^* \cdots x_k^*$.
Actually, we consider several additional generalizations,
which vary according to how we measure the size of the input,
conditions on the input, and measures of the size of the result.
For an application of the noncommutative Frobenius problem,
see Cl\'ement, Duval, Guaiana, Perrin, and 
Rindone \cite{Clement&Duval&Guaiana&Perrin&Rindone:2005}.

      In order to motivate our definitions, we consider the easiest
case first:  where $\Sigma = \lbrace {\tt 0} \rbrace$, a unary alphabet.

\section{The unary case}
\label{unary-sec}

Suppose $x_i = {\tt 0}^{a_i}$, for $1 \leq i \leq k$.  
The Frobenius problem is evidently linked to many problems over
unary languages.  It figures, for example, in estimating the size of the
smallest DFA equivalent to a given NFA \cite{Chrobak:1986}.

If $L \subseteq \Sigma^*$, by $\overline{L}$ we mean
$\Sigma^* - L$, the complement of $L$.  If $L$ is a finite language, by $|L|$ we
mean the cardinality of $L$.  Evidently we have

\begin{proposition}
     Suppose $x_i = {\tt 0}^{a_i}$
for $1 \leq i \leq k$, and write $S = \lbrace x_1, x_2, \ldots, x_k \rbrace$.
Then $S^*$ is co-finite if and only
$\gcd(a_1, a_2, \ldots, a_k) = 1$.  Furthermore, if $S^*$ is co-finite,
then the length of the longest word in $\overline{S^*}$ is 
$g(a_1, a_2, \ldots, a_k)$, and $| \, \overline{S^*} \, | = 
f(a_1, a_2, \ldots, a_k)$.
\end{proposition}

     This result suggests that one appropriate noncommutative generalization of
the condition $\gcd(a_1, a_2, \ldots, a_k) = 1$ is that
$S^* = \lbrace x_1, x_2, \ldots, x_k \rbrace^*$ be co-finite, and one
appropriate generalization of the $g$ function is the length of the longest
word not in $S^*$.

      But there are other possible generalizations.  Instead of measuring
the length of the longest omitted word, we could instead consider the
{\it state complexity} of $S^*$.  By the state complexity of a regular
language $L$, written $\stc(L)$, we mean the number of states in the
(unique) minimal deterministic finite automaton (DFA) accepting $L$.
In the unary case, this alternate measure has a nice expression in terms
of the ordinary Frobenius function:

\begin{theorem}  
Let $\gcd(a_1, a_2, \ldots, a_k) = 1$.  Then
$$ \stc( \lbrace {\tt 0}^{a_1}, {\tt 0}^{a_2}, \ldots,
{\tt 0}^{a_k} \rbrace^* ) =
	g(a_1, a_2, \ldots, a_k) + 2.$$
\end{theorem}

\begin{proof}
Since $\gcd(a_1, a_2, \ldots, a_k) = 1$, every word of length 
$> g(a_1, a_2, \ldots, a_k)$ will be in the set
$\lbrace {\tt 0}^{a_1}, {\tt 0}^{a_2}, \ldots, {\tt 0}^{a_k} \rbrace^*$.  Thus we can
accept $\lbrace {\tt 0}^{a_1}, {\tt 0}^{a_2}, \ldots, {\tt 0}^{a_k} \rbrace^*$ with a DFA
having $g(a_1, \ldots, a_k) + 2$ states, using a ``tail'' of
$g(a_1, \ldots, a_k) + 1$ states and a ``loop'' of one accepting state.
Thus $\stc( \lbrace {\tt 0}^{a_1}, {\tt 0}^{a_2}, \ldots, {\tt 0}^{a_k} \rbrace^* ) \leq
g(a_1, a_2, \ldots, a_k) + 2.$

To see $\stc( \lbrace {\tt 0}^{a_1}, {\tt 0}^{a_2}, \ldots, {\tt 0}^{a_k} \rbrace^* ) \geq
g(a_1, a_2, \ldots, a_k) + 2$, we show that the words
$$ \epsilon, {\tt 0}, {\tt 0}^2, \ldots, {\tt 0}^{g(a_1, \ldots, a_k) + 1} $$
are pairwise inequivalent under the Myhill-Nerode equivalence
relation.  Pick ${\tt 0}^i$ and ${\tt 0}^j$, $0 \leq i < j \leq g(a_1, \ldots, a_k) + 1$.
Let $L = \lbrace {\tt 0}^{a_1}, {\tt 0}^{a_2}, \ldots, {\tt 0}^{a_k} \rbrace^*$.
Choose $z = {\tt 0}^{g (a_1, \ldots, a_k) - i}$.  Then
${\tt 0}^i z = {\tt 0}^{g(a_1, \ldots, a_k)} \not\in L$, while
${\tt 0}^j z = {\tt 0}^{g(a_1, \ldots, a_k)+j-i} \in L$, since $j > i$.
\end{proof}

\begin{corollary}
Let $\gcd(a_1, \ldots, a_k) = d$.
Then
$$ \stc( \lbrace {\tt 0}^{a_1}, {\tt 0}^{a_2}, \ldots, {\tt 0}^{a_k} \rbrace^* ) =
	d (g(a_1/d, a_2/d, \ldots, a_k/d) + 1) + 1.$$
\label{statec2}
\end{corollary}

Hence it follows that
$\stc( \lbrace {\tt 0}^{a_1}, {\tt 0}^{a_2}, \ldots, {\tt 0}^{a_k} \rbrace^* ) = O(a_k^2)$.
Furthermore, this bound is essentially optimal;
since $g(n, n+1) = n^2 - n - 1$, there exist examples with
$\stc( \lbrace {\tt 0}^{a_1}, {\tt 0}^{a_2}, \ldots, {\tt 0}^{a_k} \rbrace^* ) = \Omega(a_k^2)$.

\section{The case of larger alphabets}

      We now turn to the main results of the paper.  Given as input a list of
words $x_1, x_2, \ldots, x_k$, not necessarily distinct, and
defining $S = \lbrace x_1, x_2, \ldots, x_k$, 
we can measure the size of the input in a number of different ways:

\begin{itemize}
\item[(a)] $k$, the number of words;

\item[(b)] $n = \max_{1 \leq i \leq k} |x_i|$, the length of the longest
word;

\item[(c)] $m = \sum_{1 \leq i \leq k} |x_i|$, the total number of symbols;

\item[(d)] $\sc(\lbrace x_1, x_2, \ldots, x_k \rbrace)$, the state complexity
of the language represented by the input.

\item[(e)] $\nsc(\lbrace x_1, x_2, \ldots, x_k \rbrace)$, the nondeterministic
state complexity of the language represented by the input.
\end{itemize}

     We may impose various conditions on the input:

\begin{itemize}
\item[(i)] Each $x_i$ is defined over the unary alphabet;

\item[(ii)] $S^* = \lbrace x_1, x_2, \ldots, x_k \rbrace^*$ is co-finite

\item[(iii)] $k = 2$;

\item[(iv)] $k$ is fixed.

\end{itemize}

      And finally, we can explore various measures on the size of the
result:

\begin{enumerate}

\item ${\cal L} = \max_{x \in \Sigma^* - S^*} |x|$,
the length of the longest word not in $S^*$;

\item ${\cal K} = \max_{x \in \Sigma^* - x_1^* x_2^* \cdots x_k^*} |x|$, the
length of the longest word not in $x_1^* x_2^* \cdots x_k^*$;

\item ${\cal S} = \sc (S^*)$, the state complexity of $S^*$;

\item ${\cal N} = \nsc(S^*)$, the nondeterministic state complexity of $S^*$;

\item ${\cal M} = |\Sigma^* - S^* |$, the number of words not in $S^*$;

\item ${\cal S'} = \sc(x_1^* x_2^* \cdots x_k^*)$;

\item ${\cal N'} = \nsc(x_1^* x_2^* \cdots x_k^*)$

\end{enumerate}

      Clearly not every combination results in a sensible question to study.
In order to study ${\cal L}$, the length of the longest word omitted by $S^*$.
we clearly need to impose condition (ii), that $S^*$ be co-finite.

      We now study under what conditions it makes sense to study
${\cal K} = \max_{x \in \Sigma^* - x_1^* x_2^* \cdots x_k^*} |x|$, the
length of the longest word not in $x_1^* x_2^* \cdots x_k^*$.

\begin{theorem}
     Let $x_1, x_2, \ldots, x_k \in \Sigma^+$.
Then $L = x_1^* x_2^* \cdots x_k^*$ is co-finite if and only if
$|\Sigma| = 1$ and $\gcd(|x_1|, \ldots, |x_k|) = 1$.
\end{theorem}

\begin{proof}
    If $|\Sigma| = 1$ and $\gcd(|x_1|, \ldots, |x_k|) = 1$, then a unary
word of every sufficiently long length can be attained by concatenations of the
$x_i$, so $L$ is co-finite.

      For the other direction, suppose $L$ is co-finite.  If $|\Sigma| = 1$,
let $\gcd(|x_1|, \ldots, |x_k|) = d$.  If $d > 1$, $L$ contains
only words of length divisible by $d$, and so is not-cofinite. 
So $d = 1$.

      Hence assume $|\Sigma| \geq 2$, and let $a, b$ be distinct letters
in $\Sigma$.  Let $l = \max_{1 \leq i \leq k} |x_i|$,
the length of the longest word.  Let $L' = ((a^{2l} b^{2l})^k)^+$.
Then we claim that $L' \intersect L = \emptyset$.
For if none of the $x_i$ consist of
powers of a single letter, then the longest block of consecutive identical
letters in any word in $L$ is $< 2l$, so no word in $L'$
can be in $L$.  Otherwise, say some of the $x_i$
consist of powers of a single letter.  Take any word $w$ in $L$, and count
the number $n(w)$ of maximal
blocks of $2l$ or more consecutive identical letters in
$w$.  (Here ``maximal'' means such a block is delimited on both sides
by either the beginning
or end of the word, or a different letter.)  Clearly $n(w) \leq k$.  But
$n (w') \geq 2k$ for any word in $L'$.  Thus $L$ is not co-finite, as it
omits all the words in $L'$.
\end{proof}

\section{State complexity results}

     In this section we study the measures ${\cal S} = \sc (S^*)$,
${\cal N} = \nsc(S^*)$, and
${\cal S'} = \sc(x_1^* x_2^* \cdots x_k^*)$.
     consider some results on state complexity.
First we review previous results.

     Yu, Zhuang, and Salomaa \cite{Yu&Zhuang&Salomaa:1994} showed that
if $L$ is accepted by a DFA with $n$ states, then $L^*$ can be
accepted by a DFA with at most $2^{n-1} + 2^{n-2}$ states.
Furthermore, they showed this bound is realized,
in the sense that for all $n \geq
2$, ther exists a DFA $M$ with $n$ states such that the minimal DFA
accepting $L(M)^*$ needs $2^{n-1} + 2^{n-2}$ states. This latter result
was given previously by Maslov \cite{Maslov:1970}.

    C\^{a}mpeanu, Culik, Salomaa, and Yu \cite{Campeanu&Culik&Salomaa&Yu:2001,Campeanu&Salomaa&Yu:2000} showed that if a DFA with $n$ states
accepts a {\it finite\/} language $L$, then $L^*$ can be accepted by
a DFA with at most $2^{n-3} + 2^{n-4}$ states for $n \geq 4$.  Furthermore,
this bound is actually achieved for $n > 4$ for an alphabet of size $3$ or
more.     Unlike the examples we are concerned with in this section,
however, the finite languages they construct contain exponentially many
words in $n$.

      Holzer and Kutrib \cite{Holzer&Kutrib:2003} examined the nondeterminstic
state complexity of Kleene star.  They showed that if an NFA $M$ with $n$
states accepts $L$, then $L^*$ can be accepted by an NFA with $n+1$ states,
and this bound is tight.  If $L$ is finite, then $n-1$ states suffices, and
this bound is tight.

   C\^{a}mpeanu and Ho \cite{Campeanu&Ho:2004} gave tight bounds for the
number of states required to accept a finite language whose words are
all bounded by length $n$.

\begin{proposition}
\ 
     \begin{itemize}
     \item[(a)] $\nsc( \lbrace x_1, x_2, \ldots, x_k \rbrace^* ) \leq
	  m -k + 1$.

     \item[(b)]  $\sc (\lbrace x_1, x_2, \ldots, x_k \rbrace^*) \leq
	2^{m-k+1}$.

     \item[(c)] If no $x_i$ is a prefix of any other $x_j$, then
     $\sc (\lbrace x_1, x_2, \ldots, x_k \rbrace^*) \leq m-k+2$.
     \end{itemize}
\end{proposition}

\begin{proof}

\begin{itemize}
\item[(a)]  Form an NFA from 
the trie for the words $x_1, \ldots, x_k$, sharing
a common initial state $q_0$, and having the transition on the last letter of
each word go back to $q_0$.  This NFA will have $m-k+1$ nodes.

\item[(b)]  Take the NFA from part (a) and apply the subset construction.

\item[(c)]  If no $x_i$ is a prefix of any other $x_j$, then the NFA constructed
in part (a) is actually a DFA.  One extra state is needed as a ``dead'' state.
\end{itemize}

\end{proof}

     We now consider an example providing a lower bound for the
state complexity of $\lbrace x_1, x_2, \ldots, x_k \rbrace^*$.

     Let $t$ be an integer $\geq 2$, and define words as follows:
\begin{eqnarray*}
y & := & {\tt 0} {\tt 1}^{t-1} {\tt 0} \\
x_i & := & {\tt 1}^{t-i-1} {\tt 0} {\tt 1}^{i+1}, \ \ \ \ 0 \leq i \leq t-2  \ .\\
\end{eqnarray*}
Let $S_t := \lbrace {\tt 0}, x_0, x_1, \ldots, x_{t-2}, y \rbrace$.

Thus, for example,
$$S_6 := \lbrace {\tt 0}, {\tt 1111101}, {\tt 1111011}, {\tt 1110111}, 
 {\tt 1101111}, {\tt 1011111}, {\tt 0111110} \rbrace.$$

\begin{theorem}
     $S_t^*$ has state complexity $3 t 2^{t-2} + 2^{t-1}$.
\end{theorem}

     The proof of this theorem is rather complicated, so we give
a proof of the following slightly weaker result:

\begin{theorem}
$\sc(S_t^*) \geq 2^{t-2}$.
\label{sc8}
\end{theorem}

\begin{proof}
     First, we create an NFA $M_t$ with $3t-1$ states
that accepts $S_t^*$.  This NFA has states
$$ Q = \lbrace p_0, p_1, \ldots, p_t, q_1, q_2, \ldots, q_{t-1},
	r_1, r_2, \ldots, r_{t-1} \rbrace$$
with only one final state
$F = \lbrace p_0 \rbrace$.  

For example, here is the NFA $M_6$.

\begin{figure}[H]
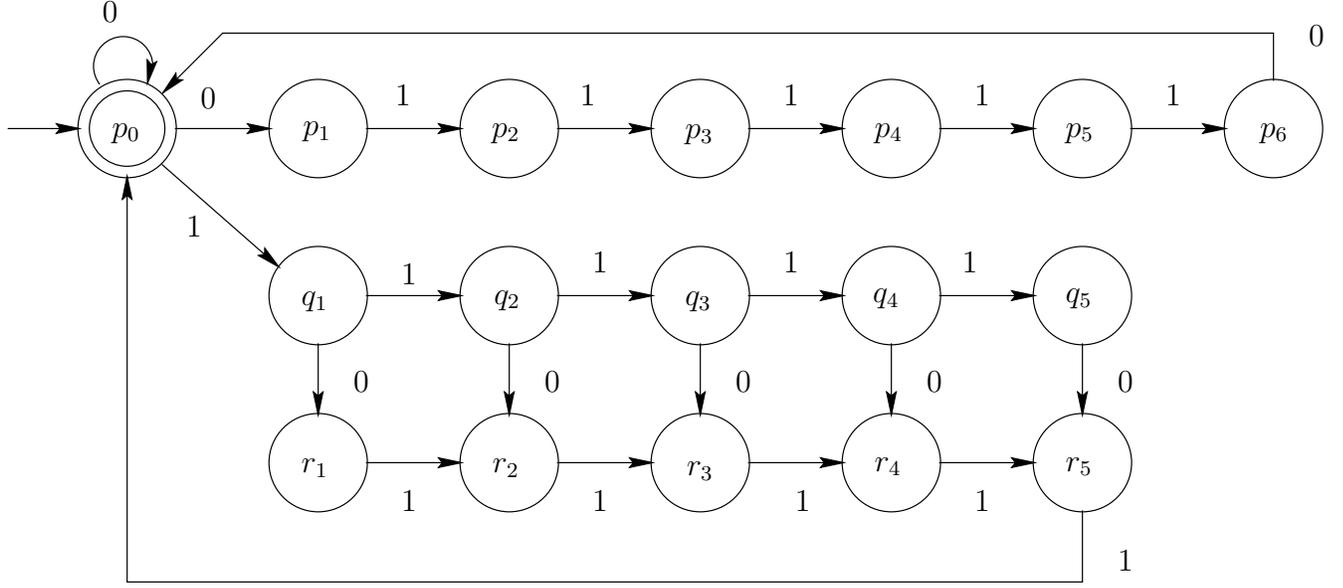

\begin{center}
\input m6.pstex_t
\end{center}
\caption{The NFA $M_6$}
\protect\label{m6}
\end{figure}

     We now determine $\delta(q, z)$ for each state $q$ of $M_t$ and
each element of $z \in S_t$.  The reader can verify that
\begin{eqnarray*}
\delta(p_0, y) &=& \lbrace r_{t-1}, p_0 \rbrace \\
\delta(p_0, x_i) &=& \lbrace p_0 \rbrace, \quad\quad 1 \leq i \leq t-2 \\
\delta(p_i, x_j) &=& \begin{cases}
	q_i, & \text{if } j = i-1; \\
	\emptyset, & \text{otherwise} .
	\end{cases} \\
\delta(p_t, y) &=& \lbrace r_{t-1} \rbrace \\
\delta(q_1, y) &=& \lbrace p_0, p_1 \rbrace \\
\delta(q_i, y) &=& \lbrace r_{i-1} \rbrace, \quad\quad 2 \leq i \leq t-1 \\
\delta(q_i, x_j) &=& \begin{cases}
	q_i, & \text{if } j \geq i; \\
	\emptyset, & \text{otherwise} . 
	\end{cases}   \quad\quad \text{for } 1 \leq i \leq t-1, 0 \leq j \leq t-2\\
\delta(r_i, y) &=& \emptyset, \quad\quad 1 \leq i \leq t-1 \\
\delta(r_i, x_j) &=& \begin{cases}
	\lbrace q_i, p_{j+1} \rbrace, & \text{if } j = i-1; \\
	\lbrace r_i \rbrace, & \text{if } j < i-1; \\
	\emptyset, & \text{otherwise} .
	\end{cases}
	\quad\quad \text{for } 1 \leq i \leq t-1, 0 \leq j \leq t-2 \\
\end{eqnarray*}

From these relations, we deduce that

\begin{eqnarray*}
\delta(\lbrace q_i, q_{i+1}, \ldots, q_{t-1}, p_t, p_0 \rbrace, y)
&=& \lbrace r_{i-1}, r_i, \ldots, r_{t-1}, p_0 \rbrace \\
\delta(\lbrace r_{i+1}, r_{i+2}, \ldots, r_{t-1}, p_0 \rbrace, x_i)
&=& \lbrace q_{i+1}, p_{i+2}, r_{i+2}, r_{i+3}, \ldots, r_{t-1}, p_0 \rbrace \\
\delta(\lbrace q_i, q_{i+1}, \ldots, q_j, p_{j+1}, r_{j+1}, r_{j+2}, 
\ldots , r_{t-1}, p_0 \rbrace, x_j) &=& \lbrace q_i, q_{i+1}, \ldots,
q_{j+1}, p_{j+2}, r_{j+2}, r_{j+3}, \ldots, r_{t-1}, p_0 \rbrace, \\
&&
\quad\quad \text{if } i \leq j \leq t-3 \\
\delta(\lbrace q_i, q_{i+1}, \ldots, q_{t-2}, p_{t-1}, r_{t-1},
 p_0 \rbrace, x_j) &=& \lbrace q_i, q_{i+1}, \ldots,
q_{t-1}, p_{t},  r_{t-1}, p_0 \rbrace \\
\end{eqnarray*}

     Let $T$ be any subset of $\lbrace r_1, r_2, \ldots, r_{t-2} \rbrace$, and
write $T = \lbrace r_{i_1}, r_{i_2}, \ldots, r_{i_j} \rbrace$ for $j$ indices
$$1 \leq i_1 < i_2 < \cdots < i_j \leq t-2 .$$
We claim that the
$2^{t-2}$ words
$$ y\  x_{t-2} y \ x_{t-3} x_{t-2} y \ x_{t-4} x_{t-3} x_{t-2} y  
 \ \cdots \  x_1 x_2 \cdots x_{t-2} y \ x_{i_1} x_{i_2} \cdots x_{i_j}  y$$
where
$$ 1 \leq i_1 < i_2 < \cdots < i_j \leq t-2,$$
are pairwise inequivalent under the Myhill-Nerode equivalence relation.

     To show this, we first argue that any subset of states of the form
$T' := \lbrace p_0, r_{t-1} \rbrace \union T$,
where $T$ is as in the previous paragraph,
is reachable from $p_0$.  From the relations above we see that
the following path reaches $T'$:
$$ \lbrace p_0 \rbrace \stackrel{y}{\longrightarrow} \lbrace p_0, r_{t-1} \rbrace \stackrel{x_{t-2}}{\longrightarrow} \lbrace q_{t-1}, p_t, p_0 \rbrace
\stackrel{ y}{\longrightarrow} \lbrace r_{t-2}, r_{t-1}, p_0 \rbrace 
\stackrel{x_{t-3}}{\longrightarrow} $$
$$\lbrace q_{t-2}, p_{t-1}, r_{t-1}, p_0
\rbrace \stackrel{x_{t-2}}{\longrightarrow} \lbrace q_{t-2}, q_{t-1}, p_t, p_0
\rbrace  \stackrel{y}{\longrightarrow}
\lbrace r_{t-3},r_{t-2}, r_{t-1}, p_0 \rbrace
\stackrel{x_{t-4}}{\longrightarrow} $$
$$ \lbrace q_{t-3}, p_{t-2}, r_{t-2}, r_{t-1},
p_0 \rbrace 
 \stackrel{x_{t-3}}{\longrightarrow} \lbrace q_{t-3}, q_{t-2}, p_{t-1}, r_{t-1}, p_0 \rbrace
\stackrel{x_{t-2}}{\longrightarrow} \lbrace q_{t-3}, q_{t-2}, q_{t-1}, p_t,
p_0 \rbrace
\stackrel{y}{\longrightarrow}$$
$$ \lbrace r_{t-4}, r_{t-3}, r_{t-2}, r_{t-1},
p_0 \rbrace \cdots  $$
$$ \stackrel{x_1 x_2 \cdots x_{t-2} y}{\longrightarrow}
\lbrace p_0, r_1, r_2, \ldots, r_{t-1} \rbrace $$
$$ 
\stackrel{x_{i_1} x_{i_2} \cdots x_{i_j} y}{\longrightarrow}
\lbrace  r_{i_1}, r_{i_2}, \ldots, r_{i_j},  r_{t-1}, p_0 \rbrace.$$

	Finally, we argue that each of these subsets of states
is inequivalent.  This is because given two distinct such subsets, say
$T'$ and $T''$, there must be an $r_i$, $1 \leq i \leq t-2$, that is
contained in one (say $T'$) but not the other.
Then reading the word $1^{t-i}$ takes $T'$ to $p_0$, but not $T''$.
\end{proof}

\begin{corollary}
     There exists a family of sets $S_t$, each consisting of $t+1$
words of length $\leq t+1$, such that $\sc(S_t^*) =
2^{\Omega(t)}$.  If $m$ is the total number of symbols in these words,
then $\sc(S_t^*) = 2^{\Omega(\sqrt{m}}$.
\end{corollary}

      Using the ideas in the previous proof, we can also create an
example achieving subexponential state complexity for
$x_1^* x_2^* \cdots x_k^*$.  

\begin{theorem}
     As before, define
\begin{eqnarray*}
y & := & {\tt 0} {\tt 1}^{t-1} {\tt 0} \\
x_i & := & {\tt 1}^{t-i-1} {\tt 0} {\tt 1}^{i+1}, \ \ \ \ 0 \leq i \leq t-2  \ .\\
\end{eqnarray*}
Let $L = ({\tt 0}^* x_1^* x_2^* \cdots x_{n-1}^* y^*)^e$ where
$e = (t+1)(t-2)/2 + 2t$.  Then $\sc(L) \geq 2^{t-2}$.
\end{theorem}

\begin{proof}
Define $A = \lbrace x_0, x_1, \ldots , x_{t-2}, y, 0 \rbrace$ and
$T = \lbrace x_1, x_2, \ldots, x_{t-2} \rbrace$.
For any subset $S$ of $T$, say $\lbrace s_1, s_2, \ldots, s_j \rbrace$
with $s_1 < s_2 < \cdots s_j$ define
$$x(S) = y x_{t-2} y x_{t-3} x_{t-2} y \cdots y x_1 x_2 \cdots x_{t-2} y
	x_{s_1} x_{s_2} \cdots x_{s_j} y .$$
Note that $x(S)$ contains $t$ copies of $y$ and at most
$(t-2)(t-1)/2 + t-2 = (t+1)(t-2)/2$ $x$'s.  Thus
$|x(S)| \leq (t+1)( t + (t+1)(t-2)/2)$ and
$|x(S)|_0 \leq  2t + (t+1)(t-2)/2 $.  

To get the bound $\sc(L) \geq 2^{t-2}$, we exhibit $2^{t-2}$ pairwise
distinct word under the Myhill-Nerode equivalence relation.
Pick two distinct subsets of $T$, say $R$ and $S$.  Since $R \not= S$,
there exists an element in one not contained in the other.  Without loss
of generality, let $m \in R$, $m \not\in S$.    By the proof of
Theorem~\ref{sc8} we have
$x(R) 1^{t-m} \in A^*$ but $x(S) 1^{t-m} \not\in A^*$.
Since $L \subseteq A^*$, $x(S) 1^{t-m} \not\in L$.  It remains
to see $x(R) 1^{t-m} \in L$.

Since $x(R) 1^{t-m} \in A^*$, there exists a factorization of
$x(R) 1^{t-m}$ in terms of elements of $A$.  However,
\begin{eqnarray*}
|x(R) 1^{t-m}| & \leq & |x(R)| + t \\
& \leq & \ (t+1)(t + (t+1)(t-2)/2 + t ) 
\end{eqnarray*}
so any factorization of
$x(R) 1^{t-m}$ into elements of $A$ contains at most $(t+1)(t-2)/2 + 2t$ copies of words
other than $0$.
Similarly
\begin{eqnarray*}
|x(R) 1^{t-m}|_0 & \leq & |x(R)| \\
&\leq & \ (t+1)(t-2)/2  + 2t
\end{eqnarray*}
so any factorization of $x(R) 1^{t-m}$ into elements of $A$
contains at most $(t+1)(t-2)/2 + 2t$ copies of the
word $0$.  
Thus a factorization of $x(R) 1^{t-m}$ into elements of $A$ is actually
contained in $L$.
\end{proof}

\begin{corollary}
     There exists an infinite family of tuples $(x_1, x_2, \ldots, x_k)$
where $m$, the total number of symbols, is $O(t^4)$, and
and $\sc(x_1^* \cdots x_k^*) = 2^{\Omega(t)}$.
\end{corollary}

      We now turn to an upper bound on the state complexity of $S^*$ in the case
where the number of words in $S$ is not specified, but we do have a bound
on the length of the longest word.

\begin{theorem}
Let $S = \lbrace x_1, x_2, \ldots, x_k \rbrace$ be a finite set
with $\max_{1 \leq i \leq k} |x_i| = n$, that is, the longest word  is
of length $n$.  Then $\sc(S^*) \leq 
{2 \over {2|\Sigma| - 1}} (2^n |\Sigma|^n - 1)$.
\label{con1}
\end{theorem}

\begin{proof}
     The idea is to create a DFA $M = (Q, \Sigma, \delta, q_0, F)$
that records the last $n-1$ symbols
seen, together with the set of the possible positions inside those
$n-1$ symbols where the factorization of the input into elements of $S$
could end.  

     Our set of states $Q$ is defined by $\lbrace [w, T] \ : \ 
|w| < n, \ S \subseteq \lbrace 0, 1, \ldots, |w| \rbrace $.  The intent
is that the DFA reaches state $[x, T]$ 
on input $y = y_1 y_2 \cdots y_i$ if and only if
$|x| = \max(n-1, i)$, $x$ is a suffix of $y$, and
$$T = \lbrace a \ :
0 \leq a \leq x {\rm \ and\ } y_1 y_2 \cdots y_{n-a} \in S^* \rbrace.$$
The initial state is $[\epsilon, \lbrace 0 \rbrace]$ and the
set of final states is $\lbrace [x, T] \ : \ 0 \in T \rbrace$.

To maintain the invariant, we define our transition function $\delta$
as follows:

If $|x| < n-1$, then $\delta([x,T], a) = [xa, U]$ 
where 
\begin{displaymath}
U = \begin{cases}
	(T+1) \union \lbrace 0 \rbrace, &
		\text{if a suffix of length $i+1$ of $xa$ is in $S$ for
		some $i \in T$;} \\
	(T+1) , & \text{otherwise.}
	\end{cases}
\end{displaymath}

If $|x| = n-1$, then $\delta([bx,T], a) = [xa, U]$ where
\begin{displaymath}
U = \begin{cases}
	((T+1) - \lbrace n \rbrace ) \union \lbrace 0 \rbrace, &
		\text{if a suffix of length $i+1$ of $bxa$ is in $S$ for
		some $i \in T$;} \\
	(T+1) - \lbrace n \rbrace, & \text{otherwise.}
	\end{cases}
\end{displaymath}

Verification that the construction works is left to the reader.  The
number of states is $\sum_{0 \leq i < n} |\Sigma|^i 2^{i+1} = 
{2 \over {2|\Sigma| - 1}} (2^n |\Sigma|^n - 1)$.
\end{proof}

\section{State complexity for two words}

      In this section we develop formulas bounding the state complexity
of $\lbrace w, x \rbrace^*$ and $w^* x^*$.  Here, as usual, 
$g(x_1, x_2)$ denotes the Frobenius function introduced in Section~\ref{intro}.

      We need the following lemma, which is of independent interest
and which generalizes a classical theorem of Fine and Wilf
\cite{Fine&Wilf:1965}.

\begin{lemma}
      Let $w$ and $x$ be nonempty words.
Let $y \in w \lbrace w,x \rbrace^\omega$ and
$z \in x \lbrace w,x \rbrace^\omega$.  Then the following conditions
are equivalent:
\begin{itemize}

\item[(a)]  $y$ and $z$ agree on a prefix of length $|w|+|x|-\gcd(|w|,|x|)$;

\item[(b)]  $wx = xw$;

\item[(c)]  $y = z$.
\end{itemize}
Furthermore, the bound in (a) is optimal, in the sense that for all
pairs of lengths $(m,n)$ there exists a pair of words $(|w|, |x|)$
such that $w^\omega$ and $x^\omega$ agree on a prefix of length
$|w|+|x|-\gcd(|w|,|x|)-1$.
\end{lemma}

\begin{proof}

(a) $\implies$ (b):
We prove the contrapositive.  Suppose $wx \not= xw$.  Without loss
of generality, we can assume $\gcd(|w|, |x|) = 1$, for if not, we group
the symbols of $w$ and $x$ into blocks of size $d = \gcd(|w|,|x|)$, obtaining
new words over a larger alphabet whose lengths are relatively prime.

Then we prove that $y$ and $z$ differ at a position $\leq |w| + |x| - 1$.
The proof is by induction on $|w|+|x|$.  

The base case is $|w|+|x|=2$.  Then $|w|=|x|=1$.  Since $wx \not=xw$, we
must have $w = a$, $x = b$ with $a \not= b$.  Then $y$ and $z$ differ at
the $1$'st position.

Now assume true for $|w|+|x| < k$.  We prove it for $|w|+|x| = k$.
If $|w|=|x|$ then $y$ and $z$ must disagree at the $|w|$'th position or earlier,
for otherwise $w = x$ and $wx = xw$, and $|w| \leq |w|+|x|-1$.
So, without loss of generality,
assume $|w| < |x|$.  If $w$ is not a prefix of $x$, then $y$ and $z$
disagree on the $|w|$'th position or earlier, and again $|w| \leq |w|+|x|-1$.

So $w$ is a proper
prefix of $x$.  Write $x = wt$ for some nonempty word $t$.
Now $wt \not= tw$, for if so, then $wx = wwt = wtw = xw$.
Then $y = w w \cdots$ and $z = wt \cdots$.  By induction
(since $|w|+|t| < k$) $w^{-1} y$ and $w^{-1} z$ disagree at position
$|w|+|t|-1$ or earlier.  Hence $y$ and $z$ diagree at position
$2|w|+|t|-1 = |w|+|x|-1$ or earlier.

\medskip

(b) $\implies$ (c):
If $wx = xw$, then by the theorem of Lyndon-Sch\"utzenberger, both $w$
and $x$ are powers of a common word $u$.  Hence $y = u^\omega = z$.

\medskip

(c) $\implies$ (a):
Trivial.

\medskip

    For the optimality statement, the words constructed in the paper
\cite{Cautis&Mignosi&Shallit&Wang&Yazdani:2003} suffice.
\label{fwcool}
\end{proof}

\begin{theorem}
     Let $w, x \in \Sigma^+$.  Then
$$
\sc( \lbrace w, x \rbrace^* ) =
	\begin{cases}
		|w|+|x|, & \text{if $wx \not= xw$}; \\
		d( g(|w|/d, |x|/d) + 1) + 2, & \text{if $wx = xw$ and
		$d = \gcd(|w|, |x|)$ }.
	\end{cases}
$$
Furthermore, this bound is tight.
\end{theorem}

\begin{proof}
      If $wx = xw$, then by a classical theorem of Lyndon and
Sch\"utzenberger \cite{Lyndon&Schutzenberger:1962}, we know there exists
a word $z$ and integers $i, j \geq 1$ such that $w = z^i$, $x = z^j$.
Thus $\lbrace w, x \rbrace^* = \lbrace z^i, z^j \rbrace^*$.  Let
$e = \gcd(i,j)$.   
Then 
$L = \lbrace z^i, z^j \rbrace^*$ consists of all words of the form
$z^{ke}$ for $k > g(i/e, j/e)$, together with some words of the form
$z^{ke}$ for $0 \leq k < g(i/e, j/e)$.  Thus, as in the proof of
Corollary~\ref{statec2}, we can accept $L$ with a ``tail'' of 
$e|z| g(i/e, j/e) + 1$ states and a ``loop'' of $e |z|$ states.
Adding an additional state as a ``dead state'' to absorb
unused transitions gives a total
of $(e |z| (g(i/e, j/e)+ 1) + 2$ states.  Since $d = e |z|$, the bound
follows.

      Otherwise, $xw \not= wx$.  Without loss of generality, let us
assume that $|w| \leq |x|$.  Suppose $w$ is not a prefix of $x$.  Let
$p$ be the longest common prefix of $w$ and $x$.  Then we can write
$w = p a w'$ and $x = p b x'$ for $a \not= b$.
Then we can accept $\lbrace w, x \rbrace^*$
with a transition diagram that has one chain of nodes labeled $p$ leading
from $q_0$ to a state $q$, and two additional chains leading from $q$ back
to $q_0$, one labled $a w'$ and one labeled $b x'$.  Since $a \not= b$, this
is a DFA.  One
additional ``dead state'' might be required to absorb transitions on letters
not mentioned.  The total number of states
is $|p| + 1 + |w'| + |x'| + 1 \leq |w| + |x|$.

      Finally, suppose $|w| \leq |x|$ and $w$ is a prefix of $x$.  
We claim it suffices to bound the longest common prefix between any word of
$w \lbrace w, x \rbrace^*$ and $x \lbrace w, x \rbrace^*$.  For if
the longest common prefix is of length $b$, we can distinguish between
them after reading $b+1$ symbols.  The $b+1$'th symbol
must be one of two possibilities, and we can use back arrows in the
transition diagram to the appropriate state.  We may need one additional
state as a ``dead state'', so the total number of states needed is $b+2$.
But from Lemma~\ref{fwcool}, we know
$b \leq |w| + |x| - 2$. 
\end{proof}

\begin{theorem}
      Let $w, x \in \Sigma^+$.  Then
$$
\sc ( w^* x^* ) =
	\begin{cases} 
		|w|+2|x|, & \text{if $wx \not= wx$}; \\
		d( g(|w|/d, |x|/d) + 1) + 2, & \text{if $wx = xw$ and
				$d = \gcd(|w|, |x|)$ }.
	\end{cases}
$$
\end{theorem}
	
\begin{proof}
    Similar to the proof of the previous theorem.  Omitted.
\end{proof}

\section{Longest word omitted}

     In this section we assume that $S = \lbrace x_1, x_2,
\ldots, x_k \rbrace$ for finite words $x_1, x_2, \ldots, x_k$, and
$S^*$ is co-finite.  We first obtain an upper bound on the length of the
longest word not in $S^*$.

\begin{theorem}
     Suppose $|x_i| \leq n$ for all $i$.  Then if $S^*$ is co-finite,
the length of the longest word not in $S^*$ is $<
{2 \over {2|\Sigma| - 1}} (2^n |\Sigma|^n - 1)$.
\label{upperbound}
\end{theorem}

\begin{proof}
    Given $S$, construct the DFA accepting
$S^*$ by the construction of Theorem~\ref{con1}.  The resulting DFA
has $q = {2 \over {2|\Sigma| - 1}} (2^n |\Sigma|^n - 1)$ states.  Now
change the ``finality'' of each state, so a final state becomes non-final
and vice versa.  This new DFA accepts $\overline{S^*}$.  Then the longest
word accepted is the length of a longest path to a final state, which
is at most $q-1$.
\end{proof}

     In the rest of this section we show that the length of the longest word
not in $S^*$ can be exponentially long in $n$.  We need several preliminary
results first.

      We say that $x$ is a {\it proper prefix} of a word $y$ if
$y = xz$ for
a nonempty word $z$.  Similarly, we say $x$ is a {\it proper suffix} of
$y$ if $y = zx$ for a nonempty word $z$.

\begin{proposition}
Let $S$ be a finite set of
nonempty words such that $S^*$ is co-finite, and $S^* \not= \Sigma^*$.
Then for all $x \in S$, there exists $x' \in S$ such that $x$ is a
proper prefix of $x$, or vice versa. Similarly, for all $x \in S$,
there exists $x' \in S$ such that $x$ is a proper suffix of $x'$, or
vice versa.
\label{prop1}
\end{proposition}

\begin{proof}
Let $x \in S$.  Since $S^* \not= \Sigma^*$,
there exists $v \in \overline{S^*}$.
Since $S^*$ is co-finite, $S^* \intersect x*v$ is nonempty.
Let $i \geq 0$ be the smallest integer such that $x^i v \in S^*$; then
$i \geq 1$, for otherwise $v \in S^*$.
Since $x^i v \in S^*$, there exist $y_1, y_2, \ldots, y_j \in S$ such
that $x^i v = y_1 y_2 \ldots y_j$.  Now $y_1 \not= x$, for otherwise by
cancelling an $x$ from both sides, we would have
$x^{i-1} v \in S^*$, contradicting the minimality of $i$.  
If $|x| < |y_1|$, then $x$ is a proper prefix of $y_1$, while if
$|x| > |y_1|$, then $y_1$ is a proper prefix of $x$.

A similar argument applies for the result about suffixes.
\end{proof}

    Next, we give two lemmas that characterize those sets $S$ such that
$S^*$ is co-finite, when
$S$ is a set $S$ containing words of no more than
two distinct lengths.

\begin{lemma}
     Suppose $S \subseteq \Sigma^m \union \Sigma^n$, $0 < m < n$,
and $S^*$ is co-finite.  Then $\Sigma^m \subseteq S$.
\label{lem2}
\end{lemma}

\begin{proof}
     If $S^* = \Sigma^*$, then $S$ must contain every word $x$ of length
$m$, for otherwise $S^*$ would omit $x$.  So assume $S^* \not= \Sigma^*$.

     Let $x \in \Sigma^m$.   Then $S^* \intersect x \Sigma^*$ is
nonempty, since $S^*$ is co-finite.   Choose $v$ such that $xv \in S^*$;
then there is a factorization
$xv = y_1 y_2 \cdots y_j$ where each $y_i \in S$.  If $y_1 \in \Sigma^m$,
then $x = y_1$ and so $x \in S$.  Otherwise $y_1 \in \Sigma^n$.   By
Proposition~\ref{prop1}, there exists $z \in S$ such that $y_1$ is a proper
prefix of $z$ or vice versa.  But since $S$ contains words of only
lengths $m$ and $n$, and $y_1 \in \Sigma^n$, we must have $z \in \Sigma^m$,
and $z$ is a prefix of $y_1$.  Then $x = z$, and so $x \in S$.
\end{proof}

\begin{lemma}
Suppose $S \subseteq \Sigma^m \union \Sigma^n$, with $0 < m < n < 2m$ and
$S^*$ is co-finite.  Then $\Sigma^l \subseteq S^*$, where
$l = m|\Sigma|^{n-m} + n-m $.
\label{lem3}
\end{lemma}

\begin{proof}
Let $x$ be a word of length $l$ that is not in $S^*$.  Then we can write
$x$ uniquely as
\begin{equation}
x = y_0 z_0 y_1 z_1 \cdots y_{|\Sigma|^{n-m} -1} z_{|\Sigma|^{n-m} - 1}
	y_{|\Sigma|^{n-m}},
\label{eq1}
\end{equation}
where $y_i \in \Sigma^{n-m}$ for $0 \leq i \leq |\Sigma|^{n-m}$,
and $z_i \in \Sigma^{2m-n}$ for $0 \leq i < |\Sigma|^{n-m}$.

Now suppose that $y_i z_i y_{i+1} \in S$ for some $i$ with
$0 \leq i < |\Sigma|^{n-m}$.
Then we can write
$$ x = \left( \prod_{0 \leq j < i} y_j z_j \right) \
y_i z_i y_{i+1} \
\left( \prod_{i+1 \leq k \leq |\Sigma|^{n-m}} z_k y_k \right).$$
Note that $|y_j z_j| = |z_k y_k| = m$.  From Lemma~\ref{lem2}, each term in
this factorization is in $S$.  Hence $x \in S^*$, a contradiction.
It follows that
\begin{equation}
y_i z_i y_{i+1} \not\in S \text{ for all $i$ with } 0 \leq i < |\Sigma|^{n-m}.
\label{ineq2}
\end{equation}

Now the factorization of $x$ in Eq.~(\ref{eq1}) uses $|\Sigma|^{n-m} + 1$
$y$'s, and there are only $|\Sigma|^{n-m}$ distinct words of length
$n-m$.  So, by the pigeonhole principle, we have $y_p = y_q$ for
some $0 \leq p < q \leq |\Sigma|^{n-m}$.  Now define
\begin{eqnarray*}
u &=& y_0 z_0 \cdots y_{p-1} z_{p-1} \\
v &=& y_p z_p \cdots y_{q-1} z_{q-1} \\
w &=& y_q z_q \cdots y_{|\Sigma|^{n-m}},
\end{eqnarray*}
so $x = uvw$.  Since $S^*$ is co-finite, there exists a smallest exponent
$k \geq 0$ such that $u v^k w \in S^*$.  

Now let $uv^k w = x_1 x_2 \cdots x_j$ be a factorization 
into elements of $S$.  Then $x_1$ is a
word of length $m$ or $n$.  If $|x_1| = n$, then comparing lengths
gives $x_1 =  y_0 z_0 y_1$.  But by (\ref{ineq2})
we know $y_0 z_1 y_1 \not\in S$.
So $|x_1| = m$, and comparing lengths gives $x_1 = y_0 z_0$.
By similar reasoning we see that $x_2 = y_1 z_1$, and so on.
Hence $x_j = y_{|\Sigma|^{n-m} - 1} z_{|\Sigma|^{n-m}-1} y_{|\Sigma|^{n-m}}
\in S$.  
But this contradicts (\ref{ineq2}).

Thus, our assumption that $x \not\in S^*$ must be false, and so $x \in S^*$.
Since $x$ was arbitrary, this proves the result.
\end{proof}

     Now we can prove an upper bound on the length of omitted words,
in the case where $S$ contains words of at most two distinct lengths.

\begin{theorem}
     Suppose $S \subseteq \Sigma^m \union \Sigma^n$, where $0 < m < n < 2m$,
and $S^* $ is co-finite.  Then
$S^* \not= \Sigma^*$, and the length of the longest word not in $S^*$ is
$\leq g(m,l) = ml - m -l $, where $l = m |\Sigma|^{n-m} + n-m$.
\label{xu5}
\end{theorem}

\begin{proof}
     Any word in $S^*$ must be a concatenation of words of length $m$ 
and $n$.  If $\gcd(m,n) = d > 1$, then $S^*$ omits all words whose length
is not congruent to $0$ (mod $d$), so $S^*$ is not co-finite, contrary
to the hypothesis.  Thus $\gcd(m,n) = 1$.  Then $S^*$ omits all words
of length $g(m,n)$, so $S^* \not= \Sigma^*$.

     By Lemmas~\ref{lem2} and \ref{lem3},
we have $\Sigma^m \union \Sigma^l \subseteq S^*$, where
$l = m |\Sigma|^{n-m} + n-m$.  Hence $S^*$ contains all words of length
$m$ and $l$; since $\gcd(m,l) = 1$, $S^*$ contains all words of length
$> g(m,l)$.
\end{proof}

\bigskip

\noindent{\it Remark.}  We can actually improve the result of the previous
theorem to arbitrary $m$ and $n$, thus giving an upper bound in the case
where $S$ consists of words of exactly two distinct lengths.  Details will
appear in a later version of the paper.

\bigskip

\begin{corollary}
Suppose $S \subseteq \Sigma^m \union \Sigma^n$, where $0 < m < n < 2m$ and
$\gcd(m,n) = 1$.  
Then $S^*$ is co-finite iff $\Sigma^m \subseteq S$ and $\Sigma^l \subseteq S^*$,
where $l = m |\Sigma|^{n-m} + n-m $.
\label{cor1}
\end{corollary}

\begin{proof}
If $S^*$ is co-finite, then by Lemmas~\ref{lem2} and \ref{lem3} we get
$\Sigma^m \subseteq S$ and $\Sigma^l \subseteq S^*$.  On the other hand,
if $\Sigma^m \subseteq S$ and $\Sigma^l \subseteq S^*$, then since
$\gcd(m,l) = 1$, every word
of length $> g(m,l)$ is contained in $S^*$, so $S^*$ is co-finite.
\end{proof}

    We need one more technical lemma.

\begin{lemma}
Suppose $S \subseteq \Sigma^m \union \Sigma^n$, where $0 < m < n < 2m$,
and $S^*$ is co-finite.  Let $\tau$ be a word not in $S^*$ where
$|\tau| = n+jm$ for some $j \geq 0$.
Then $S^* \intersect (\tau \Sigma^m )^{i-1} \tau = \emptyset$ for 
$1 \leq i < m$.
\label{tech}
\end{lemma}

\begin{proof}
     As before, since $S^*$ is co-finite we must have $\gcd(m,n) = 1$.
Define $L_i = (\tau \Sigma^m)^{i-1} \tau$ for $1 \leq i < m$.
We prove that $S^* \intersect L_i = \emptyset$ by induction on $i$.

The base case is $i = 1$.  Then $L_i = L_1 = \lbrace \tau \rbrace$.  But
$S^* \intersect \lbrace \tau \rbrace$ by
the hypothesis that $\tau \not\in S^*$.

Now suppose we have proved the result for some $i$,
$i \leq m-2$, and we want to prove it
for $i+1$.  First we show that $S^* \intersect \Sigma^{n-m} L_i = \emptyset$.
Assume that $uw \in S^*$ for some $u \in \Sigma^{n-m}$ and
$w \in L_i$.  Then there is a factorization 
\begin{equation}
uw = y_1 y_2 \cdots y_t
\label{eq3}
\end{equation}
where $y_h \in S$ for $1 \leq h \leq t$.  
Now $|uw| = n-m + (n+jm+m)(i-1) + n+jm = n(i+1) + m(ji+i-2)$.  Since
$0 < i+1 < m$, $m$ does not divide $|uw|$.   Thus at least one of the
$y_h$ is of length $n$, for otherwise (\ref{eq3}) could not be a factorization
of $uw$ into elements of $S$.  Let $r$ be the smallest index such that
$|y_r| = n$.  Then we have
$$ uw = \overbrace{y_1 y_2 \cdots y_{r-1}}^{\text{all of length $m$}}
\overbrace{y_r}^{\text{of length $n$}} y_{r+1} \cdots y_t .$$
Hence $|y_1 y_2 \cdots y_r| = m(r-1) + n = mr + n-m$.  Since, by
Lemma~\ref{lem2} we have $\Sigma^m \subseteq S$, we can write
$y_1 \cdots y_r = u z_1 \cdots z_r $, where $z_h \in S$ for $1 \leq h \leq r$.
Thus
\begin{eqnarray*}
uw &=& y_1 \cdots y_{r-1} y_r y_{r+1} \cdots y_t \\
&=& u z_1 \cdots z_r y_{r+1} \cdots y_t ;
\end{eqnarray*}
and, cancelling the $u$ on both sides, we get 
$w = z_1 \cdots z_r y_{r+1} \cdots y_t$.  But each term on the right
is in $S$, so $w \in S^*$.  But this contradicts our inductive hypothesis
that $S^* \intersect L_i = \emptyset$.

So now we know that
\begin{equation}
S^* \intersect \Sigma^{n-m} L_i = \emptyset;
\label{useful}
\end{equation} 
we'll use this fact below.

Now assume that $S^* \intersect L_{i+1} \not= \emptyset$.  Sincej
$L_{i+1} = \tau \Sigma^m L_i$, there exists
$\alpha \in \Sigma^m$ and $w \in L_i$ such that $\tau \alpha w \in S^*$.
Write $\tau \alpha w = g_1 g_2 \cdots g_p$, where $g_h \in S$ for
$1 \leq h \leq p$.  We claim that $g_h \in \Sigma^m$ for $1 \leq h \leq j+1$.
For if not, let $k$ be the smallest index such that $|g_k| = n$.  
Then by comparing lengths, we have 
$$ \overbrace{g_1 g_2 \cdots g_{k-1}}^{\text{each of length $m$}}
\overbrace{g_k}^{\text{of length $n$}} 
\overbrace{g'_1 g'_2 \cdots g'_{j-k+1}}^{\text{each of length $m$}}$$
for some $g'_1, g'_2, \ldots, g'_{j-k+1} \in \Sigma^m$.    But this
shows $\tau \in S^*$, a contradiction.
We also have $g_{j+1} \not\in \Sigma^n$, for otherwise $\tau = g_1 \cdots
g_j g_{j+1} \in S^*$, a contradiction.  

Now either $g_{j+2} \in \Sigma^m$ or $g_{j+2} \in \Sigma^n$.  In the 
former case, by comparing lengths, we see that $g_{j+3} \cdots g_p \in
\Sigma^{n-m} L_i$.  But this contradicts (\ref{useful}).   In the latter
case, by comparing lengths, we see $g_{j+3} \cdots g_p \in L_i$, 
contradicting our inductive hypothesis.  Thus our assumption that 
$S^* \intersect L_{i+1} \not= \emptyset$ was wrong, and the lemma is proved.
\end{proof}

      Now we are ready to give a class of examples achieving the bound
in Theorem~\ref{xu5}.  We define $r(n,k,l)$ to be the word of length
$l$ representing $n$ in base $k$, possibly with leading zeros.  For
example, $r(11,2,5) = {\tt 01011}$.  For integers $0 < m < n$, we
define $$T(m,n) = \lbrace r(i,|\Sigma|, n-m) {\tt 0}^{2m-n}
r(i+1,|\Sigma|,n-m) \ : \ 0 \leq i \leq |\Sigma|^{n-m} - 2 \rbrace.$$
For example, over a binary alphabet we have $T(3,5) = \lbrace {\tt
00001}, {\tt 01010}, {\tt 10011} \rbrace$.

\begin{theorem}
   Let $m, n$ be integers with $0 < m < n < 2m$ and $\gcd(m,n) = 1$,
and let $S = \Sigma^m \union \Sigma^n - T(m,n)$.  Then $S^*$ is 
co-finite and the longest words not in $S^*$ are of length
$g(m,l)$, where $l = m |\Sigma|^{n-m} + n-m $.
\label{xu6}
\end{theorem}

\begin{proof}
     First, let's prove that $S^*$ is co-finite.  Since $\Sigma^m \subseteq
S$, by Corollary~\ref{cor1} it suffices to show that $\Sigma^l \subseteq S^*$,
where $l = m |\Sigma|^{n-m} + n-m$.  

     Let $x \in \Sigma^l$, and write
$$x = y_0 z_0 y_1 z_1 \cdots y_{|\Sigma|^{n-m} - 1} z_{|\Sigma|^{n-m} -1}
	y_{|\Sigma|^{n-m}} $$
where $y_i \in \Sigma^{n-m}$ for $0 \leq i \leq |\Sigma|^{n-m}$,
and $z_i \in \Sigma^{2m-n}$ for $0 \leq i < |\Sigma|^{n-m}$.

If $y_i z_i y_{i+1} \in T(m,n)$ for all $i$, $0 \leq i < |\Sigma|^{n-m}$,
then since the base-$k$ expansions are forced to match up, we have
$y_i = r(i, |\Sigma|, n-m)$ for $0 \leq i < |\Sigma|^{n-m}$.  
But the longest such word is of length $m |\Sigma|^{n-m} + n-2m < l$,
a contradiction.  Hence $y_i z_i y_{i+1} \in S$ for some $i$.  Thus
$$ x = \left( \prod_{0 \leq j < i} y_j z_j \right) \
y_i z_i y_{i+1} \
\left( \prod_{i+1 \leq k \leq |\Sigma|^{n-m}} z_k y_k \right).$$
Note that $|y_j z_j| = |z_k y_k| = m$.   Since $\Sigma^m \subseteq S$,
this gives a factorization of $x \in S^*$.  Since $x$ was arbitrary,
we have $\Sigma^l \subseteq S^*$.

       Now we will prove that $\tau \not\in S^*$, where
$$\tau := r(0, |\Sigma|, n-m) {\tt 0}^{2m-n} r(1,|\Sigma|, n-m) {\tt 0}^{2m-n}
\cdots r(|\Sigma|^{n-m} - 1, |\Sigma|, n-m) .$$
Note that $|\tau| = |\Sigma|^{n-m} (n-m) + (|\Sigma|^{n-m} -1)(2m-n) = 
m |\Sigma|^{n-m} + n-2m = l - m$.  
Suppose there exists a factorization
$\tau = w_1 w_2 \cdots w_t$, where $w_i \in S$ for $1 \leq i \leq t$.
Since
$|\tau|$ is not divisible by $m$, at least one of these terms is of length
$n$.  Let $k$ be the smallest index such that $w_k \in \Sigma^n$.
then $\tau = w_1 \cdots w_{k-1} w_k w_{k+1} \cdots w_t$.   By comparing
lengths, we get $w_i = r(i-1, |\Sigma|, n-m) {\tt 0}^{2m-n}$ for
$1 \leq i < k$.    Thus $w_k = r(k-1, |\Sigma|, n-m) {\tt 0}^{2m-n}
r(k, |\Sigma|, n-m) \in S \intersect \Sigma^n$.  But
$r(k-1, |\Sigma|, n-m) {\tt 0}^{2m-n}
r(k, |\Sigma|, n-m) \in T(m,n)$, a contradiction.  Thus $\tau \not\in S^*$.

We may now apply Lemma~\ref{tech} to get that $S^*$ omits words
of the form $(\tau {\tt 0}^m )^{m-2} \tau$; these words are of
length $(l-m + m)(m-2) + l-m  = lm - l - m = g(m,l)$.
This completes the proof.
\end{proof}

\begin{corollary}
      For each odd integer $n \geq 5$,  there exists a set of
binary words of length at most $n$, such that
$S^*$ is co-finite and 
the longest word not in $S^*$ is of length $\Omega(n^2 2^{n/2})$.
\end{corollary}

\begin{proof}
   Choose $m = (n+1)/2$ and apply Theorem~\ref{xu6}.
\end{proof}

\begin{example}
    Let $m = 3$, $n = 5$, $\Sigma = \lbrace {\tt 0, 1} \rbrace$.
Then  $S = \Sigma^3 + \Sigma^5 - \lbrace {\tt 00001}, {\tt 01010},
{\tt 10011} \rbrace$.  Then a longest word not in $S^*$ is
${\tt 00001010011 \, 000 \, 00001010011 }$, of length $25$.
\end{example}

\section{Number of omitted words}

Recall that $f(x_1, x_2, \ldots, x_k)$ is the classical function which,
for positive integers $x_1, \ldots, x_k$ with $\gcd(x_1, \ldots, x_k) = 1$,
counts the number of integers not representable as a non-negative integer
linear combination of the $x_i$.  In this section we consider a 
generalization of this function to the setting of a free monoid, replacing
the integers $x_i$ with finite words in $\Sigma^*$, and replacing the condition
$\gcd(x_1, \ldots, x_k) = 1$ with the requirement that
$\lbrace x_1, \ldots, x_k \rbrace^*$ be co-finite.

We have already studied this in the case of a unary alphabet in
Section~\ref{unary-sec}, so let us assume that $\Sigma$ has at least
two letters.

\begin{theorem}
     Let $x_1, x_2, \ldots, x_k \in \Sigma^*$ be such that $|x_i| \leq n$ for
$1 \leq  i \leq n$.  Let $S = \lbrace x_1, x_2, \ldots, x_k  \rbrace$
and suppose $S^*$ is co-finite.  Then 
$${\cal M} = 
| \Sigma^* - S^* | \leq  {{|\Sigma|^q - 1} \over {|\Sigma| - 1}},$$
where $q = {2 \over {2 |\Sigma| - 1}} (2^n |\Sigma|^n - 1)$.
\end{theorem}

\begin{proof}
From Theorem~\ref{upperbound}, we know that if $S^*$ is co-finite, the length of
the longest omitted word is $< q$,
where $q = {2 \over {2 |\Sigma| - 1}} (2^n |\Sigma|^n - 1)$.  The total
number of words $< q$ is
$1 + |\Sigma| + \cdots + |\Sigma|^{q-1} =
{{|\Sigma|^q - 1} \over{|\Sigma| - 1}}.$
\end{proof}

     We now give an example achieving a doubly-exponential lower bound
on $\cal M$.

\begin{theorem}
   Let $m, n$ be integers with $0 < m < n < 2m$ and $\gcd(m,n) = 1$,
and let $S = \Sigma^m \union \Sigma^n - T(m,n)$, where $T$ was introduced
in the previous section.    Then $S^*$ is co-finite and $S^*$ omits
at least $2^{|\Sigma|^{n-m}} - |\Sigma|^{n-m} - 1$ words.
\end{theorem}

\begin{proof}
       Similar to that of Theorem~\ref{xu6}.
\end{proof}

\section{Conclusion}

     We have generalized the classical Frobenius problem on integers to the
noncommutative setting of a free monoid.  Many problems remain, including
improving the upper and lower bounds presented here, and examining the
computational complexity of the associated decision problems.  We will
examine these problems in a future paper.

\end{document}